\documentclass[twocolumn,showpacs,aps,prl,superscriptaddress]{revtex4}

\newcommand{\DE}{\ensuremath{\Delta E}\xspace}
\newcommand{\mES}{\ensuremath{m_\mathrm{ES}}\xspace}

\usepackage{graphicx}
\usepackage{dcolumn}
\usepackage{amsmath}
\usepackage{epsfig}
\usepackage{bigstrut}

\input babarsym.tex

\newcommand{\BABARPubYear}    {06}
\newcommand{\BABARPubNumber}  {046}
\newcommand{\SLACPubNumber} {12031}

\newcommand{\BFa} {2.55 \pm 0.05 \pm 0.16}
\newcommand{\BFb} {2.79 \pm 0.08 \pm 0.17}
\newcommand{\BFc} {4.90 \pm 0.07 \pm 0.22}
\newcommand{\BFd} {5.52 \pm 0.17 \pm 0.42}
\newcommand{\cosdeltaA} {0.872^{+0.008+0.031}_{-0.007-0.029}}
\newcommand{\cosdeltaB} {0.924^{+0.019+0.063}_{-0.017-0.054}}
\newcommand{\cosdeltaCLA} {99.9\%~}
\newcommand{\cosdeltaCLB} {85.7\%~}
\newcommand{\deltaA} {{29.2^{\circ}}^{+0.8^{\circ}+3.3^{\circ}}_{-0.9^{\circ}-3.8^{\circ}}}
\newcommand{\deltaB} {{22.5^{\circ}}^{+2.4^{\circ}+6.1^{\circ}}_{-3.1^{\circ}-9.9^{\circ}}}
\newcommand{\AratioA} {0.655^{+0.015+0.042}_{-0.014-0.042}}
\newcommand{\AratioB} {0.624^{+0.027+0.065}_{-0.026-0.063}}

\def\figurebox#1#2#3{%
    \def\arg{#3}%
    \ifx\arg\empty
    {\hfill\vbox{\hsize#2\hrule\hbox to #2{\vrule\hfill\vbox to #1{\hsize#2\vfill}\vrule}\hrule}\hfill}%
    \else
    {\hfill\epsfbox{#3}\hfill}%
    \fi}

\begin{document}

\preprint{\babar-PUB-\BABARPubYear/\BABARPubNumber} 
\preprint{SLAC-PUB-\SLACPubNumber} 

\begin{flushleft}
\babar-PUB-\BABARPubYear/\BABARPubNumber\\
SLAC-PUB-\SLACPubNumber\\
\end{flushleft}

\title{
{\large \boldmath
Branching fraction measurement of $\Bzb \ra D^{(*)+} \pim$
and $\Bm \ra D^{(*)0} \pim$ \\   
and isospin analysis of $\Bbar \ra D^{(*)}\pi$ decays} 
}


%
\author{B.~Aubert}
\author{R.~Barate}
\author{M.~Bona}
\author{D.~Boutigny}
\author{F.~Couderc}
\author{Y.~Karyotakis}
\author{J.~P.~Lees}
\author{V.~Poireau}
\author{V.~Tisserand}
\author{A.~Zghiche}
\affiliation{Laboratoire de Physique des Particules, F-74941 Annecy-le-Vieux, France }
% ...
\collaboration{The \babar\ Collaboration}
\noaffiliation

\date{\today}

\begin{abstract}
Using 65 million $\Upsilon(4S)\ra\BB$ events collected with the 
\babar\ detector at the PEP-II \epem storage ring
at the Stanford Linear Accelerator Center,
we measure the color-favored branching fractions  
${\cal B}(\Bzb \ra \Dp \pim) = (\BFa)\times 10^{-3}$,
${\cal B}(\Bzb \ra \Dstarp \pim) = (\BFb)\times 10^{-3}$,
${\cal B}(\Bm \ra \Dz \pim) = (\BFc)\times 10^{-3}$ and
${\cal B}(\Bm \ra \Dstarz \pim) = (\BFd)\times 10^{-3}$,
where the first error is statistical and the second is systematic.
With these results and the current world average
for the branching fraction for the color-suppressed decay
$\Bzb \ra D^{(*)0}\pi^{0}$, the cosines of the strong phase
difference $\delta$ between the $I=1/2$ and $I=3/2$ isospin amplitudes
are determined to be 
$ \cos \delta = \cosdeltaA$
for the $\Bbar \ra D\pi$ process and 
$ \cos \delta = \cosdeltaB$
for the $\Bbar \ra D^{*}\pi$ process. Under the isospin symmetry,  
the results for $\cos \delta$ suggest the presence of final-state 
interactions in the $D\pi$ system.

\end{abstract}

\pacs{13.25.Hw, 12.15.Hh, 11.30.Er}

\maketitle


The $\Bbar \ra D\pi$ and $\Bbar \ra D^{*}\pi$ processes provide very good 
opportunities to test the theories of hadronic $B$-meson decays due to
their clean and dominant hadronic decay channels.
With the development of heavy quark effective theory (HQET)
\cite{HQET1, HQET2} and soft collinear effective theory (SCET) 
\cite{SCET1, SCET2}, the theoretical description for these hadronic 
decays has improved considerably, and the factorization hypothesis in 
heavy quark hadronic decay has been put on a more solid basis.
The three decay amplitudes {$\cal{A}$} for $\Bbar \ra D\pi$ 
can be expressed in terms of two isospin amplitudes, 
$A_{1/2}$ and $A_{3/2}$, under the isospin symmetry of the strong interaction:
\begin{equation}
{\cal{A}}(\Bbar^{0} \ra D^{+}\pi^{-}) 
 =  \sqrt{1/3}A_{3/2} + \sqrt{2/3}A_{1/2},
\end{equation}
\begin{equation}
\sqrt{2}{\cal{A}}(\Bbar^{0} \ra D^{0}\pi^{0}) 
 =  \sqrt{4/3}A_{3/2} - \sqrt{2/3}A_{1/2},
\end{equation}
\begin{equation}
{\cal{A}}(B^{-} \ra D^{0}\pi^{-}) 
 =  \sqrt{3}A_{3/2},
\end{equation}
where isospin amplitudes $A_{1/2}$ and $A_{3/2}$ correspond to the transitions 
into $D\pi$ final states with pure $I=1/2$ and $I=3/2$ isospin eigenstates  
\cite{Neubert:2001, Beneke:2000}. An identical decomposition 
holds for $\Bbar \ra D^{*}\pi$ decays. The isospin amplitudes are not 
necessarily the same in the $\Bbar \ra D\pi$ and $\Bbar \ra D^{*}\pi$
systems. In the context of QCD factorization \cite{Beneke:2000}, 
$A_{1/2}$ and $A_{3/2}$ for $\Bbar \ra D\pi$ (similarly for 
$\Bbar \ra D^{*}\pi$) are related by
\begin{equation}
\frac{A_{1/2}}{\sqrt{2} A_{3/2}} = 1 + O(\Lambda_{\rm QCD}/m_{b}),
\end{equation}
where $m_{b}$ is the $b$-quark mass and $\Lambda_{\rm QCD}$ is the QCD scale.
The deviation of the ratio $A_{1/2}/(\sqrt{2} A_{3/2})$ from unity is a 
measure of the departure from the heavy-quark limit. 
The QCD factorization implies that the relative phase $\delta$ of $A_{1/2}$ 
and $A_{3/2}$ is $O(\Lambda_{\rm QCD}/m_{b})$. 
Final-state interactions (FSI) in the $I=3/2$ and $I=1/2$ channels 
can lead to a non-zero $\delta$. 
A large value of $\delta$ will substantially suppress the destructive 
interference for the color-suppressed decay $\Bzb \ra D^{(*)0}\pi^{0}$, 
thereby increasing the associated branching fraction. 

Recent experimental results on the color-suppressed decay
$\Bzb \ra D^{(*)0}\pi^{0}$ \cite{CLEO:2002a, Belle:2002, Babar:2004} provide 
evidence for a sizable relative strong interaction 
phase between color-favored and color-suppressed $\Bzb \ra D^{(*)} \pi$ 
decay amplitudes.
It has been suggested \cite{Neubert:2001} that improved
measurements of the color-favored hadronic two-body decay of the $B$ meson 
will lead to a better understanding of these QCD effects. 
Further experimental results on the color-favored decay $\Bbar \ra D\pi$
suggest the presence of final-state interactions in the $\Bbar \ra D\pi$
process \cite{CLEO:2002b}. This paper presents new measurements of the 
branching fractions of
$B^{-} \ra D^{(*)0}\pi^{-}$ and $\Bzb \ra D^{(*)+}\pi^{-}$
(charge conjugation is implied throughout this paper) 
and of the relative phase $\delta$. 


This analysis uses $(65.2\pm 0.7)\times 10^{6}$ $B\Bbar$ pairs collected at
the $\Upsilon(4S)$ resonance with the  \babar\ detector~\cite{babar:nim} 
at the PEP-II asymmetric-energy storage ring during the 2001-2002 data 
taking period.
Charged tracks are detected by a 5-layer silicon
vertex tracker and a 40-layer drift chamber. Hadrons are identified
by measuring the ionization energy loss ${\rm d}E/{\rm d}x$ in the tracking 
system and the opening angle of the Cherenkov radiation in a ring-imaging 
detector. Photons are identified by an electromagnetic calorimeter.
These systems are mounted inside a 1.5-T solenoidal superconducting magnet.

Kaon and pion candidates are selected from charged-particle tracks 
using ${\rm d}E/{\rm d}x$ and the Cherenkov light signature.
Each charged track, except the track used as the soft pion to
reconstruct $\Dstarp \ra \Dz \pip$, is required to have at least 
12 hits in the drift chamber and a transverse momentum greater 
than 100~\mevc. \Dz and \Dp candidates are reconstructed in the $K^{-}\pi^{+}$
and $K^{-}\pi^{+}\pi^{+}$ channels, respectively. In each case, $D$ meson 
candidates are required to have a mass within 3$\sigma$ of the mean 
reconstructed mass value, where the mass resolution  $\sigma$
is approximately 7~\mevcc for \Dz and 6~\mevcc for \Dp. 
A vertex fit is performed on \Dz (\Dp) candidates with the 
mass constrained to the nominal value \cite{PDG:2006}. 
A \Dz candidate is combined with a low momentum \pip or \piz to
form a \Dstarp or \Dstarz candidate, where the \piz candidate 
is formed from two photon candidates and must have an invariant mass 
between 120 and 145~\mevcc. 
Combinations with an invariant mass difference $\Delta m = m_{\Dz\pi}-m_{\Dz}$ 
between 143 and 148~\mevcc for \Dstarp and between 138 and 146~\mevcc 
for \Dstarz, corresponding to $\pm 3\sigma$ about the $\Delta m$ peak, 
are retained.
Each $B$ meson candidate is reconstructed using the selected $D$ or \Dstar  
candidate and an additional charged track that is not consistent with the kaon 
hypothesis. 

To reject jet-like continuum background events, the normalized second 
Fox-Wolfram moment $R_2$~\cite{FW}, computed with charged tracks and neutral 
clusters, is required to be less than 0.5. We also require $|\cos\theta_T|$ to 
be less than $0.85$, where $\theta_T$ is the angle between the thrust axis of 
the $B$ candidate and the thrust axis of the rest of the event in the \epem 
center-of-mass (CM) frame. 

$B$ candidates are identified using the beam-energy-substituted mass 
$\mES= \sqrt{(\sqrt{s}/2)^2 - p^{*2}}$ and energy difference 
$\DE= E^*- \sqrt{s}/2$, where $E^*$ and $p^*$ are the energy and momentum 
of the reconstructed \B candidate and $\sqrt{s}$ is the total energy in 
the \epem CM frame. $B$ signal candidates have $\mES \sim m_{B}$,
the $B$ meson mass, and $\DE \simeq 0$, within their respective resolutions.  
The resolution in \DE, $\sigma_{\Delta E}$, for various $B$ modes
ranges from 15.7 to 18.1 MeV.
We require that $|\DE - \langle \DE \rangle |< 3\sigma_{\Delta E}$.
For events with more than one 
$B$ candidate, a $\chi^{2}$ is defined with the $D$ mass $m_{D}$, 
$\Delta m$ and their resolutions as  
\begin{equation}
 \chi^{2} =\left(\frac{m_{D} - \langle m_{D} \rangle }
          {\sigma_{m_{D}}}\right)^{2} +
          \left(\frac{
          \Delta m-\langle \Delta m\rangle}
         {\sigma_{\Delta m}}\right)^{2}
\end{equation}
and the candidate with the smallest $\chi^{2}$ is chosen. 

\begin{figure*}[htb]
\vspace{0.1cm}
\begin{center}
\includegraphics[width=0.7\textwidth]{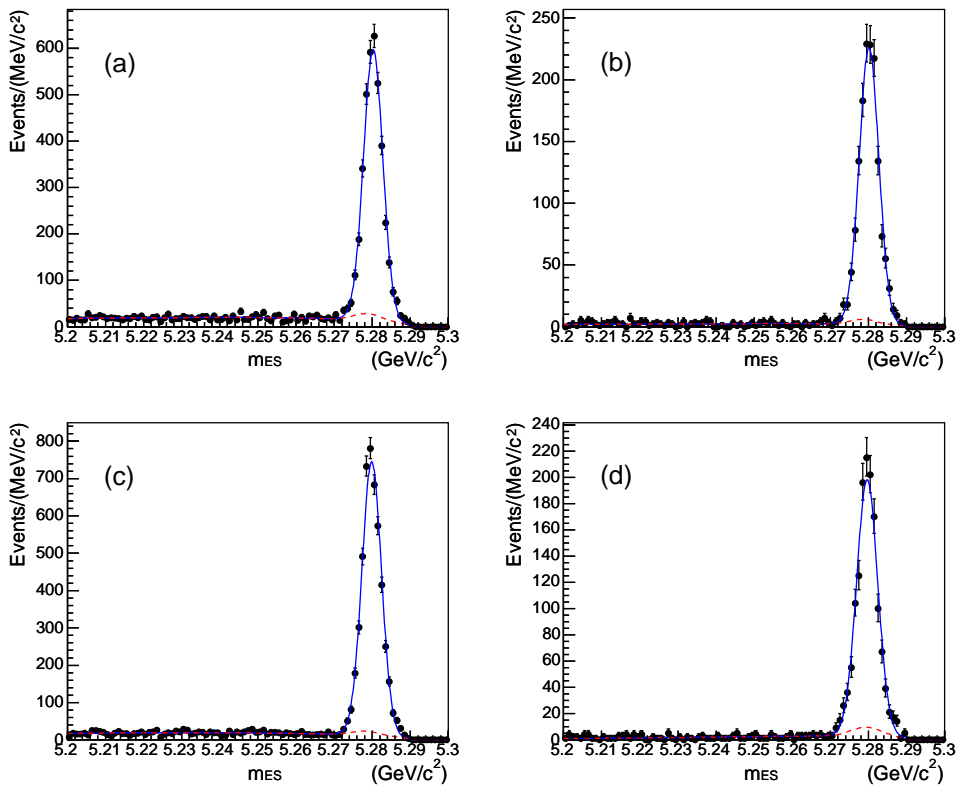}
\end{center}
\caption{
Fit of $\mES$ distributions for the $B\ra D^{(*)}\pi$ candidates in data:
             (a) $\Bzb \ra D^{+}\pi^{-}$,
             (b) $\Bzb \ra D^{*+}\pi^{-}$,
             (c) $B^{-}\ra D^{0}\pi^{-}$,
             (d) $B^{-}\ra D^{*0}\pi^{-}$.
The fit is shown as a solid line and is described in the text.
The background component (including peaking background) is shown as a
dashed line.
}
\label{fig:xfit_mes}
\end{figure*}

The event yield $n$ for each mode of $\Bbar \ra D^{(*)} \pi^{-}$ is 
extracted by fitting the \mES distribution of the selected $B$ candidates 
with an unbinned extended maximum likelihood fit. The \mES distribution 
is fit to the sum of a signal component, modeled as a Gaussian, and a 
background shape. The background shape is parameterized as the sum of a 
Gaussian, representing the peaking background events that peak in \mES, and 
a phase space parameterization function \cite{argus} representing 
non-peaking combinatorial background and continuum events. 
The parameters describing the background shape, including the 
relative normalization of the peaking component, are determined by fitting 
Monte Carlo (MC) simulated samples, with the signal 
events removed. The total signal and background event yields, as well as the 
shape parameters describing signal events, are free parameters in the fit. 
The fitted \mES distributions for each of the $B$ meson decay modes are
presented in Fig.~\ref{fig:xfit_mes}. The peaking background yield 
$n_{\rm pb}$ is about (2--4)\%
of the observed $B$ signal yield, as shown in Table \ref{table-d1pi-bf}.

For each studied $B$ decay mode of $\Bbar \ra D^{(*)}\pi$,
the branching fraction is calculated as:
\begin{equation}
 {\cal{B}} (\Bbar \ra D^{(*)}\pi) 
= \frac{n}{2f N_{B\Bbar} \varepsilon  
{\cal{B}}(D^{(*)})}. 
\end{equation}
Here $N_{B\Bbar}$ is the total number of $B\Bbar$ pairs;
$\varepsilon$ is the efficiency determined from signal Monte Carlo events;
$f$ represents $f_{+-}$ or $f_{00}$, the charged or neutral $B$ meson 
production ratios at the $\Upsilon(4S)$, which we assume to be 
$f_{+-} = f_{00} = 0.5$; and
${\cal{B}}(D^{(*)})$ is the branching fraction
of $D$ or $D^{*}$ decaying to its reconstructed final state \cite{PDG:2006}. 
The branching fractions we obtain are reported in Table \ref{table-d1pi-bf}.

\begin{table*}[htb]
\renewcommand{\arraystretch}{1.25}
\begin{center}
\caption{
\label{table-d1pi-bf}
Yield of signal ($n$) and peaking background ($n_{\rm pb}$), 
efficiency ($\varepsilon$), and branching fraction (${\cal{B}}$) for 
each $\Bbar \ra D^{(*)}\pi$ decay mode.}
\begin{tabular*}{0.69\textwidth}{lcccc} \hline \hline
~~~~Mode  &~~~~~$n$ &~~~~~$n_{\rm pb}$ &~~~~~$\varepsilon$ (\%) 
&~~${\cal{B}}$ ($\times 10^{-3}$)  \\ \hline 
~~$\Bzb\ra D^{+}\pi^{-}$ 
&~~~~~ 3593 $\pm$ 63 
&~~~~~ 114 $\pm$ 14 
&~~~~~ 22.8 $\pm$ 0.2  
&~~~~~ $\BFa$ \\ 
~~$\Bzb \ra D^{*+}\pi^{-}$ 
&~~~~~ 1411 $\pm$ 39 
&~~~~~ 28   $\pm$ 6
&~~~~~ 30.2 $\pm$ 0.2  
&~~~~~ $\BFb$ \\ 
~~$B^{-} \ra D^{0}\pi^{-}$ 
&~~~~~ 4606 $\pm$ 70 
&~~~~~ 89   $\pm$ 14
&~~~~~ 37.9 $\pm$ 0.2  
&~~~~~ $\BFc$ \\ 
~~$B^{-} \ra D^{*0}\pi^{-}$ 
&~~~~~ 1297 $\pm$ 39 
&~~~~~ 51   $\pm$ 8
&~~~~~ 15.5 $\pm$ 0.1  
&~~~~~ $\BFd$ \\ \hline
\hline
\end{tabular*}
\end{center}
\end{table*}

The final states $D^{(*)}\pi$ selected by this analysis are, in general,
accompanied by some small amount of final state radiation (FSR). We
model final state radiation in our experiment with PHOTOS \cite{PHOTOS1},
which predicts that 6-7\% of our selected events, varying slightly with
decay mode, are accompanied by an average FSR energy of about 17 MeV.
Approximately two-thirds of this energy is produced in the initial $B$
decay, while the remainder is generated in the $D^{(*)}$ decay.

We summarize systematic uncertainties on the measurements from 
various sources in Table \ref{table-d1pi-sys}. 
$\Delta N_{B\Bbar}$ is the uncertainty on the total number of $B\Bbar$ 
pairs in data. The error on the efficicency, $\Delta \varepsilon$, is due to 
signal Monte Carlo sample statistics. 
The uncertainty from combinatoric background
is estimated as the difference in the $B$ yields obtained when fixing and 
floating the non-peaking background parameters in the \mES fit.
The uncertainty from peaking background is estimated as the $B$ yield 
change by varying the peaking background parameters and the ratio
of peaking background to non-peaking background within their errors in 
the \mES fit.
The uncertainties due to the differences in $D^{(*)}$ masses and $\Delta E$
between data and Monte Carlo samples are estimated by comparing the 
efficiencies using their resolutions and means from data and Monte Carlo 
samples in the event selection. 
The uncertainty due to $D$ vertexing is estimated by comparing vertexing 
perfomance in data and Monte Carlo samples.
The uncertainties in tracking, particle identification, and $\pi^{0}$ 
reconstruction efficiencies are due to potential residual inaccuracies 
in the Monte Carlo simulation, after correcting for known differences. 
The dominant uncertainty is from the $D^{(*)}$ branching fractions 
${\cal{B}}(D^{(*)})$ and the tracking efficiency.

\begin{table*}[htb]
\renewcommand{\arraystretch}{1.25}
\begin{center}
\caption{
\label{table-d1pi-sys}
Relative systematic errors in the branching fractions of 
$\Bbar \ra D^{(*)}\pi$ decays from different sources.}
\begin{tabular*}{0.78\textwidth}{lcccc} \hline \hline
~~Systematic error & ~~$\Bzb \ra D^{+}\pi^{-}$ & ~~$\Bzb \ra D^{*+}\pi^{-}$
            & ~~$B^{-}\ra D^{0}\pi^{-}$ & ~~$B^{-}\ra D^{*0}\pi^{-}$ \\ \hline
~~$\Delta N_{B\Bbar}$                 & 1.1\% & 1.1\% & 1.1\% & 1.1\% \\
~~${\cal{B}}(D^{(*)})$                & 3.6\% & 2.0\% & 1.8\% & 5.0\% \\
~~$\Delta f$                          & 1.6\% & 1.6\% & 1.6\% & 1.6\% \\
~~$\Delta \varepsilon$                & 1.0\% & 0.5\% & 0.5\% & 0.7\% \\
~~Non-peaking background shape        & 2.8\% & 0.5\% & 1.9\% & 1.3\% \\
~~Peaking background shape            & 0.4\% & 0.4\% & 0.3\% & 0.6\% \\
~~Data/MC difference of $m_{D}$, $\Delta m$
                                      & 0.2\% & 1.3\% & 0.4\% & 2.9\% \\
~~Data/MC difference of $\Delta E$    & 0.5\% & 0.2\% & 0.6\% & 0.7\% \\
~~$D^{-}$ and $D^{0}$ vertexing       & 0.2\% & 0.1\% & 0.1\% & 0.1\% \\
~~Particle identification efficiency  & 2.0\% & 2.0\% & 1.5\% & 1.5\% \\
~~Tracking efficiency                 & 3.2\% & 4.9\% & 2.4\% & 2.4\% \\
~~$\pi^{0}$ reconstruction efficiency &  -    &  -    &  -    & 3.0\% \\
\hline
~~Total                               & 6.3\%& 6.2\%& 4.4\%& 7.6\%\\
\hline \hline
\end{tabular*}
\end{center}
\end{table*}

With the branching fractions of the four color-favored decay modes
$\Bzb \ra D^{(*)+}\pi^{-}$ and $B^{-} \ra D^{(*)0}\pi^{-}$,
as well as the two color-suppressed modes 
$\Bzb \ra D^{(*)0}\pi^{0}$, one can calculate $\cos \delta$. 
Following Ref.~\cite{rosner:1999} (equations have been modified to use the 
notation from Ref.~\cite{Neubert:2001}),
$\cos \delta$ for $\Bbar \ra D\pi$ (similarly for $\Bbar \ra \Dstar\pi$) 
can be expressed as       
\begin{equation}
\cos\delta = \frac{3\Gamma(D^{+}\pi^{-}) + \Gamma(D^{0}\pi^{-})
-6\Gamma(D^{0}\pi^{0})}{6\sqrt{2}|A_{1/2}A_{3/2}|},
\end{equation}
\begin{equation}
|A_{3/2}|^{2} = \frac{1}{3}\Gamma(D^{0}\pi^{-}),
\end{equation}
\begin{equation}
|A_{1/2}|^{2} = \Gamma(D^{+}\pi^{-})
+\Gamma(D^{0}\pi^{0})
-\frac{1}{3}\Gamma(D^{0}\pi^{-}).
\end{equation}
Using the measured branching fractions in this analysis, the ratio of the 
$B$ lifetimes $\tau(B^{-})/\tau(\Bzb) = 1.071 \pm 0.009$ \cite{PDG:2006}, 
and the branching fractions 
${\cal{B}}(\Bzb \ra D^{0}\pi^{0}) = (0.291 \pm 0.028)\times 10^{-4}$ 
and ${\cal{B}}(\Bzb \ra D^{*0}\pi^{0}) = (0.27 \pm 0.05)\times 10^{-4}$ 
\cite{PDG:2006}, we calculate $\cos\delta$ and 
$|A_{1/2}/(\sqrt{2}A_{3/2})|$ for $\Bbar \ra D\pi$ and $\Bbar \ra \Dstar\pi$
decays.

\begin{figure}[htb]
\begin{center}
 \includegraphics[width=0.50\textwidth]{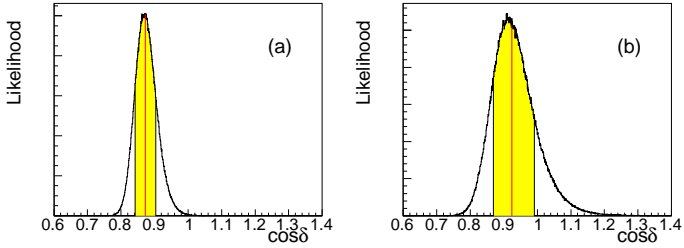}
\end{center}
\caption{
Likelihood function (arbitrary unit in vertical axis) of 
$\cos \delta$ obtained from the 
ensemble of $10^{6}$ Monte Carlo experiments described in the text
for process (a) $\Bbar \ra D\pi$ and (b) $\Bbar \ra D^{*}\pi$. 
The shaded area in the plots is 68.27\% of the total area. 
}
\label{Fig-xphys-d1pi-delta}
\end{figure}

\begin{figure}[htb]
\begin{center}
 \includegraphics[width=0.50\textwidth]{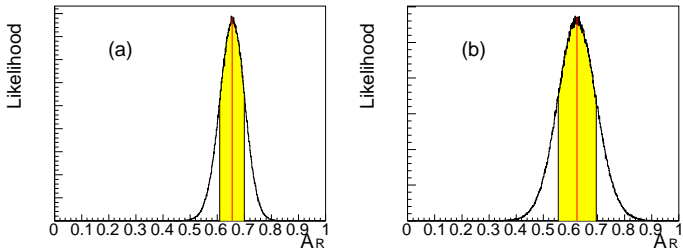}
\end{center}
\caption{
Likelihood function (arbitrary unit in vertical axis) of $A_{R} \equiv 
|A_{1/2}/\sqrt{2}A_{3/2}|$ obtained 
from the ensemble of $10^{6}$ Monte Carlo experiments described in the text
for processes (a) $\Bbar \ra D\pi$ and (b) $\Bbar \ra D^{*}\pi$. 
The shaded area in the plots is 68.27\% of the total area. 
}
\label{Fig-xphys-d1pi-A12A32} 
\end{figure}

To estimate the systematic error on $\cos\delta$ 
for $\Bbar \ra D\pi$ (and, similarly, $\Bbar \ra D^{*}\pi$), we use a 
Monte Carlo technique \cite{CLEO:2002b}.  
We simulate $10^{6}$ experiments, varying the measured branching 
fractions, the used color-suppressed decay branching fraction, and 
$\tau(B^{-})/\tau(\Bzb)$ about their central values according to   
Gaussian distributions where their errors are taken as the sigmas of the
Gaussian distributions, to calculate the $\cos\delta$.
The correlation of the systematic errors between the two color-favored 
decay modes in the $\cos\delta$ calculation is taken into account. 
We assume the errors are uncorrelated between the color-favored 
and color-suppressed modes. The statistical error on $\cos\delta$
is estimated in a similar fashion, with only the statistical errors
on the branching fractions of color-favored modes are used in the procedure.  
The resulting normalized distribution of $\cos\delta$, 
ie., the estimated likelihood function of $\cos\delta$, is obtained.
Figure \ref{Fig-xphys-d1pi-delta} shows the likelihood function 
of $\cos\delta$ from the 
described experiments in which both the statistical and systematic errors 
are taken into account. 
 
We define $\pm 1 \sigma$ confidence interval of $\cos\delta$ as the 
integral of its likelihood function over the region around the nominal 
value of $\cos\delta$, which is calculated from the central values of the 
branching fractions, to 68.27\% (half below and half above the 
nominal value) of the total area. The results are
\begin{equation}
\cos \delta = \cosdeltaA 
\end{equation}
for the $\Bbar \ra D\pi$ system and 
\begin{equation}
\cos \delta = \cosdeltaB
\end{equation}
for the $\Bbar \ra D^{*}\pi$ system, 
where the first error is statistical and the second is systematic. 
These results correspond to 
$|\delta| = \deltaA $ and
$|\delta| = \deltaB $, for the $\Bbar \ra D\pi$ system and the 
$\Bbar \ra D^{*}\pi$ system, respectively.  
By comparing the likelihood function integral of $\cos\delta$ in 
region [0,1] with the full range integral, we exclude $\cos\delta \geq 1$ 
at a probability of \cosdeltaCLA for the $\Bbar \ra D\pi$ system and 
\cosdeltaCLB for the $\Bbar \ra D^{*}\pi$ system.

Similarly, we obtain 
\begin{equation}
 \left|\frac{A_{1/2}}{\sqrt{2}A_{3/2}}\right| = \AratioA
\end{equation}
and 
\begin{equation}
 \left|\frac{A_{1/2}}{\sqrt{2}A_{3/2}}\right| = \AratioB
\end{equation}
for the $\Bbar \ra D\pi$ and $\Bbar \ra D^{*}\pi$ system, respectively,
where the first error is statistical and the second is systematic.
The likelihood function from the simulated experiments, with both 
statistical and systematic errors are taken into account, 
is shown in Fig.~\ref{Fig-xphys-d1pi-A12A32}.

In summary, we have measured the branching fractions for the color-favored 
$\Bzb \ra D^{(*)+} \pim$ and $\Bm \ra D^{(*)0} \pim$ decays. 
Using these measurements together with the current world averages
for ${\cal{B}}(\Bzb \ra D^{0}\pi^{0})$ 
and ${\cal{B}}(\Bzb \ra D^{*0}\pi^{0})$, we extract the cosines of the 
relative strong phase $\delta$ in the $D\pi$ and $D^{*}\pi$ systems, 
and the ratios of the $I=3/2$ and $I=1/2$ isospin amplitudes. 
Our results for the $\Bbar \ra D^{(*)}\pi$ branching fractions,
except for $B^{-} \ra D^{*0}\pi^{-}$,  are consistent
with the current world average values \cite{PDG:2006} but have a better 
precision. The branching fraction of $B^{-} \ra D^{*0}\pi^{-}$ from this 
measurement is greater than the world average by about 2$\sigma$. 
Our results for $\cos \delta$ differ from unity  
by about 4.3$\sigma$ 
for $\Bbar \ra D\pi$ decays and 1.1$\sigma$ for 
$\Bbar \ra D^{*}\pi$ decays. The result of $\cos \delta$ for 
$\Bbar \ra D\pi$ decays is consistent with the result in Refs.
\cite{Babar:2004, CLEO:2002b}, and under the isospin symmetry it 
suggests the presence of final-state interactions in $\Bbar \ra D\pi$ decays.

We are grateful for the excellent luminosity and machine conditions
provided by our \pep2\ colleagues, 
and for the substantial dedicated effort from
the computing organizations that support \babar.
The collaborating institutions wish to thank 
SLAC for its support and kind hospitality. 
This work is supported by
DOE
and NSF (USA),
NSERC (Canada),
IHEP (China),
CEA and
CNRS-IN2P3
(France),
BMBF and DFG
(Germany),
INFN (Italy),
FOM (The Netherlands),
NFR (Norway),
MIST (Russia),
MEC (Spain), and
PPARC (United Kingdom). 
Individuals have received support from the
Marie Curie EIF (European Union) and
the A.~P.~Sloan Foundation.


\end{document}